\documentclass[sigconf,screen,nonacm]{acmart}
\setcopyright{none}
\usepackage{amsmath}
\usepackage{booktabs}
\usepackage{tabularx}
\usepackage{graphicx}
\usepackage{pgfplots}
\pgfplotsset{compat=1.18}
\usetikzlibrary{patterns} %
\usepackage{placeins}
\usepackage[normalem]{ulem} %
\newcommand{\liella}{LIELLA}

\newcommand{\BEST}{\textsc{best}}

\newcommand{\OFF}{\textsc{off}}
\newcolumntype{Y}{>{\raggedright\arraybackslash}X}

\hypersetup{pdfauthor={Yuexun Huang, Kfir Sulimany, Ke Lin, Xiangyu Ren, Dirk Englund, Liang Jiang},pdfsubject={Research preprint},pdfkeywords={photonic computing, personalization, forward-only adaptation}}
\makeatletter
\@ifundefined{hyxmp@parse@acmart}{}{\let\hyxmp@parse@acmart\relax}
\makeatother
\AddToHook{cmd/maketitle/after}{\hypersetup{pdfsubject={Research preprint}}}

\begin{document}
\title{LIELLA: Private Edge Adaptation with Photonic Weight Streams}
\author{Yuexun Huang}
\affiliation{\institution{The University of Chicago}\city{Chicago}\country{USA}}
\email{yesunhuang@uchicago.edu}
\author{Kfir Sulimany}
\affiliation{\institution{MIT}\city{Cambridge}\country{USA}}
\email{kfir@mit.edu}
\author{Ke Lin}
\affiliation{\institution{William \& Mary}\city{Williamsburg}\country{USA}}
\email{leonard.keilin@gmail.com}
\author{Xiangyu Ren*}
\affiliation{\institution{The University of Edinburgh}\city{Edinburgh}\country{United Kingdom}}
\email{xiyu.ren@outlook.com}
\author{Dirk Englund*}
\affiliation{\institution{MIT}\city{Cambridge}\country{USA}}
\email{englund@mit.edu}
\author{Liang Jiang*}
\affiliation{\institution{University of Chicago}\city{Chicago}\country{USA}}
\email{liangjiang@uchicago.edu}
\renewcommand{\shortauthors}{Huang et al.}

\begin{abstract}
Although photonic accelerators have shown promise for energy-efficient machine learning inference, privately personalizing large models across multiple edge clients remains difficult under tight memory, precision, and hardware constraints. We present LIELLA, a weight-streaming architecture that enables multiple edge clients to personalize a shared model through compact private adapters. The two proposed photonic adapter architectures accommodate both backpropagation and forward-only optimization while substantially reducing modeled client-side electrical energy consumption compared with electronic adapters. Under simulated hardware noise, clients sharing the same streamed backbone achieve nearly clean accuracy on QNLI and SST-2, even with attojoule-scale optical energy per MAC at each receiver. The architecture also enables hardware-aware adaptation among clients performing the same task, allowing their individual private adapters to achieve further accuracy improvements. For a gigabyte-scale TinyLlama backbone, LIELLA maintains a megabyte-scale inference memory footprint and kilobyte-scale persistent adapter storage per client, while achieving strong performance through forward-only adaptation. These findings further highlight hardware-driven opportunities to develop smaller yet more effective parameter-efficient fine-tuning (PEFT) methods and more accurate, noise-robust forward-only optimization algorithms.
\end{abstract}
\maketitle

\section{Introduction}
Fine-tuning large models for different users and tasks is increasingly in demand across various applications such as personal assistants, recommendation systems, image generation, and healthcare~\cite{tan2024oppu,dang2025personalizedDiffusion,li2026mta,zhu2026personalizedMedVlm}. Deploying large personalized models on resource-constrained edge devices already poses substantial memory and computational challenges, while enabling private, on-device fine-tuning imposes even more stringent resource demands~\cite{xu2025resourceSurvey,gao2025mobizo}. Photonic computing offers a promising path toward energy-efficient inference and training~\cite{shastri2021photonics,sludds2022netcast,chen2026optovit}, but extending these capabilities to private fine-tuning across multiple edge clients remains an architectural challenge. Parameter-efficient fine-tuning (PEFT) reduces the number of trainable parameters~\cite{han2024peftSurvey,hu2022lora}, yet conventional local execution still requires storing and repeatedly accessing the much larger frozen backbone~\cite{xu2025resourceSurvey}, which limits its benefits in addressing the broader memory and computational challenges of edge deployment. Offloading fine-tuning to remote infrastructure can alleviate these resource constraints, but transferring sensitive training data or client-specific model updates introduces significant privacy risks~\cite{elsheikh2026privacySurvey,zhu2019deepLeakage}.

\noindent\textbf{Weight-streaming computing.}
Weight streaming enables a new paradigm for distributed inference, allowing multiple edge clients to share a centrally hosted model backbone with low client-side memory and energy demands. This paradigm is particularly well suited to photonic systems, where energy-efficient photoelectric multiplication and wavelength-division multiplexing (WDM) naturally support shared weight broadcasting~\cite{hamerly2019photoelectric,hamerly2021wdmBroadcast}. Netcast experimentally demonstrated weight-streaming inference over an 86-km deployed fiber link and accurate image recognition with receiver-side optical energy as low as 40 aJ per MAC (Multiply–Accumulate)~\cite{sludds2022netcast,hamerly2022netcast}. AIR-LLM extends the weight-broadcast principle to radio-frequency LLM inference~\cite{gao2025wise,vadlamani2025miwen,gao2026airllm}. These systems establish a promising paradigm for sharing a model backbone across multiple edge clients. However, transforming this paradigm from inference to private personalization remains an open architectural challenge, as each client must independently fine-tune the shared model while keeping its training data and adapter state private.

\noindent\textbf{Challenges in private fine-tuning.}
Supporting independent private fine-tuning over a shared weight stream introduces three architectural challenges. 
First, each client must be able to independently train its own personalized model using the shared backbone while keeping its training data, gradients, and adapter state local. 
Second, the architecture must support different optimization methods for evaluating their effectiveness, identifying viable fine-tuning strategies, and characterizing trade-offs in training accuracy, energy, and hardware requirements. For example, backpropagation (BP) requires transposed backbone operators and intermediate activation storage or recomputation, whereas zeroth-order (ZO) optimization avoids these dependencies at the cost of additional forward evaluations and potentially noisier gradient estimates~ \cite{malladi2023mezo,zhang2024zoBenchmark, kim2026meftSurvey,gao2025mobizo}. 
Third, private adapter architectures must be designed to enable effective fine-tuning with low client-side memory and energy costs, without imposing substantial overhead on the shared weight-streaming service. This requires joint consideration of adapter parameterization, electronic and photonic execution, private-state storage and movement, hardware noise and precision, and server-side requirements. Addressing these challenges requires an end-to-end co-design that integrates independent private adaptation, effective optimization methods, and hardware-aware adapter with a shared weight-streaming service.

\noindent\textbf{Prior Works.}

Netcast demonstrates optical weight streaming, AIR-LLM explores radio-frequency weight broadcasting, and QAMNet studies efficient optical encoding for neural computation~ \cite{sludds2022netcast,hamerly2022netcast, gao2025wise,vadlamani2025miwen,gao2026airllm,bacvanski2024qamnet}. However, these approaches primarily target inference and do not address on-device training. Photonic training architectures, including FICONN, L2ight, Mirage, and Opto-ViT-v2, explore in-situ learning, high-precision training, and noise-resilient fine-tuning~ \cite{bandyopadhyay2024ficonn,gu2021l2ight, demirkiran2024mirage,chen2026optovit}. Beyond photonics, AHWA-LoRA applies low-rank adapters to hardware-aware fine-tuning on analog in-memory computing~\cite{li2026ahwalora}. Similarly, MobiZO and MeBP improve the efficiency of digital edge fine-tuning through zeroth-order optimization and memory-efficient backpropagation~ \cite{gao2025mobizo,song2025mebp}. Nevertheless, these works mainly focus on local model training within individual computing devices rather than on distributed private adaptation across multiple clients sharing the same backbone service. Petals and DLoRA offload backbone computation to remote servers while retaining client-specific adaptation~ \cite{borzunov2023petals,gao2024dlora}. This exposes data-dependent intermediate representations to the server and requires backbone computation for each client's inputs, posing serious privacy risks and limiting server-side amortization. Furthermore, none of them exploits the co-design with private adapter hardware.

\noindent\textbf{LIELLA architecture.}
To enable private fine-tuning over a shared model, we introduce \emph{Live-streaming Independent Edge Learning with Lightweight Adaptation} (\liella), a distributed fine-tuning architecture that allows multiple edge clients to independently personalize a shared weight-streaming backbone. 
Building upon Netcast \cite{sludds2022netcast}, a remote server streams the frozen backbone weights as optical signals, while each client modulates its local inputs and integrates the received signals to perform the corresponding Multiply-Accumulate. LIELLA then extends this inference pipeline by incorporating lightweight private adapters into the client-side computation. Each client evaluates its own training loss and updates its private adapter locally, without transmitting its training data or revealing their adapter parameters to the server.
Multiple clients can thus independently train and deploy their own compact adapters as the \emph{soul} of the personalized model for different tasks and hardware conditions, using the same backbone services as the \emph{body}. 
\uline{Overall, our primary goal is to enable private personalization for multiple edge clients while minimizing edge memory and energy costs and maintaining robustness against hardware noise.}

\textbf{Firstly, we propose two designs for private photonic adaptation} based on the shared Netcast weight stream. Netcast Integrated Core Optics (NICO) performs all client-private adapter computation optically at the receiver using the streaming optical signal. Mixed Adapter Kernel Integration (MAKI) executes the fixed adapter operations optically while retaining client-private
adapter computation electronically. We treat Electronic Local Integration (ELI) as a
baseline, which retains the optical backbone stream but performs all adapter computation electronically at the client, then maps LoRA-FA~\cite{zhang2023lorafa} (FA) and
LoRA-XS~\cite{balazy2024loraxs} (XS) onto these three designs to support different adapter
parameterizations within the same weight-streaming system.

\textbf{Secondly, we develop a shared optical weight-streaming execution framework} that supports inference and private fine-tuning across multiple edge clients. Building upon the low-noise server/simple-client (LN/S) signaling scheme \cite{sludds2022netcast,hamerly2022netcast}, we design a noise-aware scaling method to transform the frozen digital backbone into the shared optical stream. Specifically, we select input scaling from task-independent public calibration data to minimize the discrepancy between noisy optical and digital inference and derive output scaling to satisfy the receiver dynamic-range constraints for both forward and transposed operations. The resulting shared service supports inference and forward-only (FO) fine-tuning through forward weight streams and extends to backpropagation (BP) through additional transposed weight streams.

\textbf{Thirdly, we develop a physics-aware photonic simulator and resource models} to evaluate private adaptation over shared optical weight streams. Our simulator models key hardware nonidealities across different adapter designs and execution modes. We also model client-side memory and electrical energy consumption using workload-specific accounting and component parameters based on published hardware measurements and literature. 
Together, these allow us to evaluate resource-efficient private fine-tuning across clients performing different tasks over a shared weight stream, as well as hardware-aware personalization among clients performing the same task under heterogeneous receiver conditions. This further enables systematic comparisons of different PEFT parameterizations and optimization methods in terms of accuracy, noise robustness, and resource efficiency.

\textbf{Lastly, we evaluate our framework across various tasks and hardware conditions.} With simulated receiver noise of 640 aJ per optical MAC, clients sharing a frozen backbone achieve over 85\% accuracy on QNLI and over 90\% on SST-2. Hardware-aware private adapters further improve accuracy by up to 11 percentage points for clients performing the same task under varying receiver conditions. With 9,856 trainable parameters (38.5 KiB) and 4.81 KiB of frozen client-private rank calibration, XS achieves 87.7\% SST-2 accuracy through forward-only optimization under noiseless photonic simulation, with 7.46 MiB of modeled client inference memory. Under nominal component assumptions, NICO with XS reduces modeled client-side electrical energy by more than 70\% versus ELI for the evaluated forward-only training workload.

Our main contributions are:
\begin{itemize}
    \item We enable independent private model adaptation across multiple clients sharing a frozen optical stream, without cumulative server-side backbone computation or heavy client-side storage.
    \item We propose two photonic private-adapter designs that distribute fixed and client-private computations across optical and electronic paths, and compare them with a pure electronic baseline. 
    \item We establish a forward-only and transpose-enabled fine-tuning service protocol for ZO and BP, then evaluate their learning and resource trade-offs under simulated photonic noise, including client memory, modeled electrical energy.
    \item To the best of our knowledge, we are the first to systematically investigate how different PEFT parameterizations and ZO methods affect private fine-tuning on distributed photonic weight-streaming hardware, revealing new incentives for smaller, more effective adapters and accurate, query-efficient, noise-robust ZO algorithms.
\end{itemize}

\section{Background and Motivation}
\label{sec:background}

\subsection{Optical Weight Streaming}

Netcast implements optical weight streaming through
photonic matrix-vector multiplication (MVM)
~\cite{sludds2022netcast,hamerly2022netcast}.
For a weight matrix $W\in\mathbb{R}^{M\times N}$,
the server encodes $w_{mn}$ onto wavelength
$\lambda_m$ at time step $n$, broadcasting the
weight matrix over $M$ wavelength channels and
$N$ time steps.
At the client, a broadband Mach--Zehnder
modulator (MZM) encodes each input $x_n$ through
$x_n=\cos(2\theta_n)$, simultaneously modulating
all wavelength channels.
After wavelength demultiplexing, balanced
photodetectors integrate the resulting signals
over time to compute the MVM.

In the low-noise-server/simple-client (LN/S)
configuration~\cite{hamerly2022netcast},
the server rescales the weight magnitude in
optical power and its sign in the differential
detection channel. To accommodate the finite dynamic range of optical
receivers, a digital weight matrix
$W\in\mathbb{R}^{M\times N}$ is mapped
to an optically encoded matrix $\bar W$ through
positive diagonal scaling matrices $D_{\mathrm{in}}$
and $D_{\mathrm{out}}$:
\begin{equation}
\bar W=D_{\mathrm{out}}^{-1}WD_{\mathrm{in}},
\qquad
\bar x=\frac{D_{\mathrm{in}}^{-1}x}{s},
\end{equation}
where $s=\|D_{\mathrm{in}}^{-1}x\|_\infty$.
The scaling matrices are public and shared across
clients, while $s$ is computed locally for each
input. The expected photoelectron
counts at the two detector channels are then calculated as,
\begin{equation}
\mu_{mn}^{\pm}
=\mathbb{E}[K_{mn}^{\pm}]
=\frac{n_0|\bar w_{mn}|}{2}
\left[1\pm\operatorname{sgn}(\bar w_{mn})
\bar x_n\right].
\label{eq:netcast_counts}
\end{equation}
with $K_{mn}^{\pm}$ denoting the detected photoelectron
counts for the $\pm$ differential channel, respectively. 
The integrated charges are therefore
$Q_m^\pm=e\sum_{n=1}^{N}K_{mn}^\pm$,
which gives the balanced photodetection counts
\begin{equation}
\begin{aligned}
\frac{\mathbb{E}[Q_m^+-Q_m^-]}{en_0}
&=\frac{1}{n_0}\sum_{n=1}^{N}
(\mu_{mn}^+-\mu_{mn}^-)
=\sum_{n=1}^{N}\bar w_{mn}\bar x_n
=(\bar W\bar x)_m.
\end{aligned}
\label{eq:netcast_mac}
\end{equation}
Here, $n_0$ is the calibrated electron
count for a unit-magnitude weight, with $e$
being the elementary charge.
The original digital result is thus recovered as
$Wx=sD_{\mathrm{out}}(\bar W\bar x)$.
Under LN/S signaling, the total expected detected
photoelectron count per symbol is
$\mu_{mn}^++\mu_{mn}^-=n_0|\bar w_{mn}|$.
Consequently, the choice of scaling matrices
directly affects shot noise and receiver dynamic
range, making optical scaling a critical design
consideration rather than a purely numerical
normalization.

Importantly, LIELLA aims to provide a shared optical
service in which multiple clients independently
fine-tune distinct private adapters while consuming
the same streamed backbone. The server shall not perform client-specific encoding or
computations, nor introduce additional MAC
accumulations for individual adapters, as doing
so would undermine the computational reuse
of optical broadcasting. All client-specific inputs,
adapter computations, and signal accumulation
shall instead remain at the receivers. This
requires a common optical weight representation
and shared scaling matrices across clients,
despite their potentially different tasks, input
distributions, and hardware noise conditions.
These constraints motivate a task- and
client-independent encoding and computing strategy that balances
optical noise and dynamic-range requirements
without increasing the server-side burden.

\subsection{Parameter-Efficient Fine-Tuning}

Parameter-efficient fine-tuning (PEFT) enables
the adaptation of large pretrained models by optimizing
a compact set of parameters while keeping the
backbone frozen, substantially reducing trainable states and optimization overhead compared with
full fine-tuning. In low-rank adaptation (LoRA), each adapted
weight matrix $W$ is replaced by $W+\Delta W$,
where $\Delta W$ is a low-rank perturbation
~\cite{hu2022lora}.  In this work, we consider two low-rank PEFT
methods with distinct trainable parameter
structures,
\begin{equation}
\begin{aligned}
\text{LoRA-FA:}\quad
\Delta W &= BA,\\
\text{LoRA-XS:}\quad
\Delta W &= S_{\mathrm{out}}CS_{\mathrm{in}},
\end{aligned}
\label{eq:adapter_parameterization}
\end{equation}
where LoRA-FA freezes $A$ and trains $B$
~\cite{zhang2023lorafa}, whereas LoRA-XS
constructs fixed factors
$S_{\mathrm{out}}=U_r\Sigma_r$ and
$S_{\mathrm{in}}=V_r^{\mathsf T}$ from the
rank-$r$ truncated SVD of the pretrained
weight matrix $W$, and trains only the
tiny core $C$~\cite{balazy2024loraxs}.  For $L$ adapted linear layers, the corresponding numbers
of trainable private parameters are
\begin{equation}
\begin{aligned}
P_{\mathrm{FA}}
&=r_{\mathrm{FA}}\sum_{\ell=1}^{L}
d_{\mathrm{out},\ell},\\
P_{\mathrm{XS}}
&=Lr_{\mathrm{XS}}^2,
\end{aligned}
\label{eq:adapter_private_parameters}
\end{equation}
where $r_{\mathrm{FA}}$ and $r_{\mathrm{XS}}$ denote the respective ranks and $d_{\mathrm{out},\ell}$ is the output dimension of module $\ell$. For our evaluated backbone, 1.1 B TinyLlama configuration, with 22 Transformer blocks and seven adapted linear layers per block ($L=154$), rank-16 LoRA-FA requires 6,307,840 private parameters (24.06 MiB at FP32), while rank-8 LoRA-XS requires only 9,856 (38.5 KiB), reducing persistent private parameter storage by 640$\times$ compared with LoRA-FA.

\begin{figure}[t]
\centering
\includegraphics[width=\columnwidth]{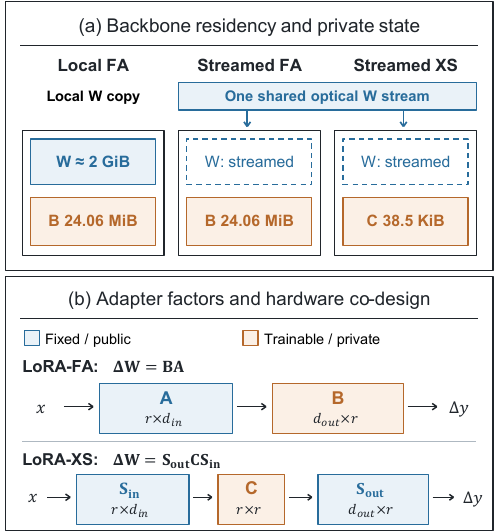}
\caption{Weight streaming and adapter--hardware co-design. (a) Streaming eliminates client-resident backbone copies ($\approx$2 GiB in FP16). Private FP32 trainable weights occupy 24.06 MiB for FA and 38.5 KiB for XS. (b) Their distinct fixed and trainable factors enable different adapter design. }
\label{fig:motivation}
\end{figure}

However, reducing trainable parameters alone does not actually produce a significant memory advantage if storing the frozen backbone is necessary. As shown in Fig.~\ref{fig:motivation}(a), for the 1.1 B model in FP16,
the backbone weights alone already occupy about 2 GiB, dominating the size of the private adapter. Consequently, increasingly compact PEFT methods provide marginal improvement in on-device model storage as long as the backbone remains resident. Optical weight streaming eliminates local backbone storage, but each client must still execute its private adapter. The distinct fixed and trainable factors of LoRA-FA and LoRA-XS, as shown in Fig.~\ref{fig:motivation}(b), create opportunities to reduce memory-access and energy costs for realizing ultra-compact, energy-efficient private adaptation at the edge through optical-electronic hardware co-design. Fixed transformations may be executed optically following the same pipeline as the streaming weight, while client-private operations can be performed either optically at the receiver or electronically, with different implications for memory access, modulation, and energy consumption. 

\subsection{Photonic Optimization}

Fine-tuning private adapters requires repeated
optimization over client-side private parameters
$\theta$. Forward-only (FO) zeroth-order
optimization estimates gradients using only
loss evaluations, without backpropagation
through the model~\cite{malladi2023mezo,gao2025mobizo}.
A typical two-point gradient estimator is
\begin{equation}
\hat{g}_{\mathrm{FO}}
=
\frac{\mathcal{L}(\theta+\epsilon u)
-\mathcal{L}(\theta-\epsilon u)}
{2\epsilon}u,
\label{eq:fo_estimator}
\end{equation}
where $u$ is a random perturbation direction
and $\epsilon$ is the perturbation magnitude.
FO can therefore reuse an inference-compatible
optical weight stream without requiring
backward propagation or its associated
activation storage. However, estimating
gradients requires repeated forward evaluations,
increasing the number of optical computation
requests. Recent methods, including SubZero
and SubZero+, employ low-dimensional
perturbations to improve optimization
efficiency~\cite{yu2025subzero,yu2026subzeroplus,
zhang2024zoBenchmark}.

In contrast, the familiar backpropagation (BP) computes
gradients through the network using the
chain rule. For a linear layer $y=Wx$,
the backward operation can be calculated as
\begin{equation}
\frac{\partial\mathcal{L}}{\partial x}
=
W^{\mathsf T}
\frac{\partial\mathcal{L}}{\partial y},
\label{eq:bp_transpose}
\end{equation}
which requires the transposed weight operators. Consequently, optical BP requires both forward and transpose-oriented weight streams, as well as retained activations or recomputation~\cite{song2025mebp,kim2026meftSurvey}. Compared with FO, BP avoids repeated zeroth-order gradient queries but introduces additional backward-service and memory requirements.

These differences become particularly important under photonic hardware noise. Finite optical energy introduces shot noise, while electronic readout and detector dynamic-range constraints further limit computation accuracy~\cite{hamerly2019photoelectric, sludds2022netcast,shastri2021photonics}. For FO, noise in repeated loss evaluations propagates into their finite differences, increasing gradient-estimation variance, especially at small $\epsilon$.
For BP, optical errors affect both forward activations and backward gradient propagation. Thus, the robustness and effectiveness of an optimization method depend not only on its noiseless convergence but also on its optical query requirements, receiver noise sensitivity,
and client-side memory footprint. These challenges motivate a systematic investigation of optimization methods under shared photonic hardware constraints, particularly query-efficient and noise-robust FO methods.

\section{LIELLA Architecture}
\label{sec:contract}

\subsection{Shared Optical Service}
LIELLA separates a public, immutable backbone service from independently owned client adaptation. For client $i$ and an adapted linear layer $\ell$, the forward computation is carried out as
\begin{equation}
y_{i,\ell}=W_\ell x_{i,\ell}+\Delta W_{i,\ell}x_{i,\ell},
\label{eq:client_layer}
\end{equation}
where $W_\ell$ is the public, frozen backbone weight shared by all clients, while the input $x_{i,\ell}$ and adapter perturbation $\Delta W_{i,\ell}$ are kept private by each client. 
The server broadcasts the same optically encoded weight stream $W_\ell$ to all clients. Each client modulates the received optical signals with its local input $x_{i,\ell}$, and performs balanced photodetection and temporal integration to obtain an estimate of $W_\ell x_{i,\ell}$. Each receiver combines its backbone result with its own adapter correction and executes the remaining data-dependent operations locally. 

\begin{figure*}[t]
\centering
\includegraphics[width=\textwidth]{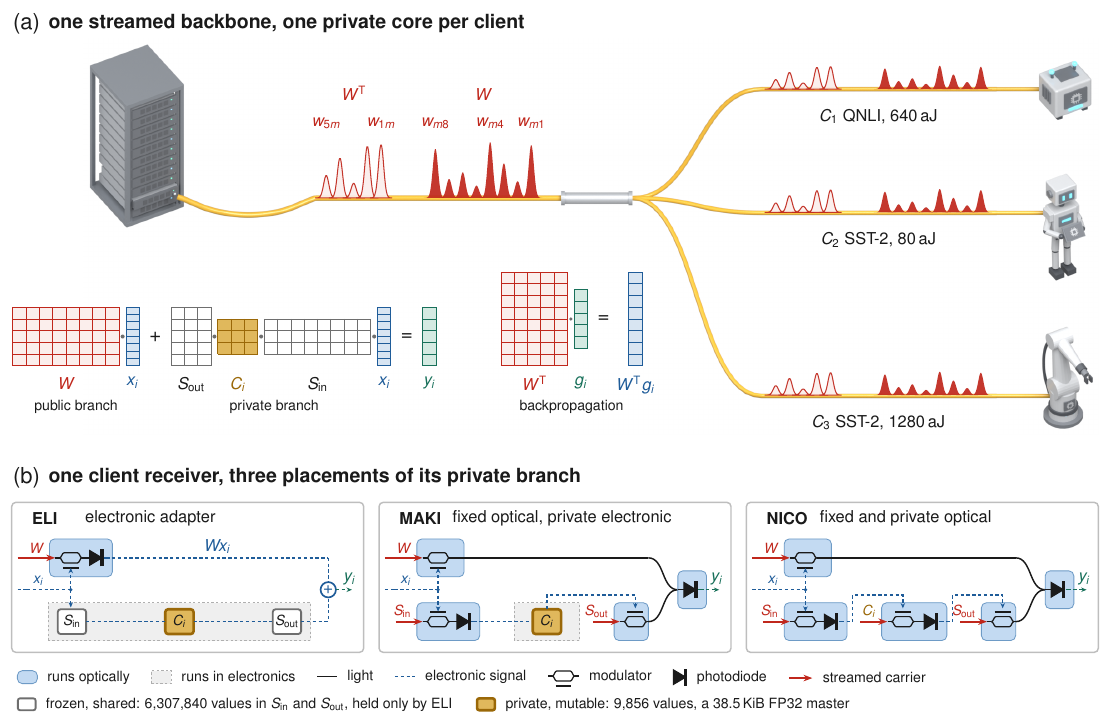}
\caption{Illustration of LIELLA's service architecture and the private-adapter design.
(a) An identical public weight stream supports QNLI~\cite{wang2018glue}
and SST-2~\cite{socher2013sst} clients under different receiver
conditions without server-side personalized computation.
(b) For the illustrated LoRA-XS adapter, ELI executes all adapter factors
electronically, MAKI maps fixed factors onto the optical path,
and NICO additionally executes the private core on the receiver-side
optical path. }
\label{fig:architecture}
\end{figure*}

The overall architecture of LIELLA is shown in Figure~\ref{fig:architecture} (a), which decouples a shared optical backbone service from client-specific private adaptation through receiver-side optical-electronic integration. A single optical weight stream serves multiple independently adapting clients. In this illustrative example, one client is fine-tuning for QNLI, while two others are performing adaptation for SST-2 under different receiver conditions. They consume the same public weight stream, but independently perform input modulation, photodetection, and private adapter updates. Each client retains its inputs, labels, losses, private adapter parameters and optimizer state, and performs adaptation locally. This enables client-specific personalization without additional server-side backbone computation or leaking their private data.

\subsection{Photonic Adapter Designs}
Figure~\ref{fig:architecture}(b) illustrates the two
photonic adapter architectures, Netcast Integrated
Core Optics (NICO) and Mixed Adapter Kernel
Integration (MAKI), compared with Electronic Local
Integration (ELI) as a baseline. They share the
public backbone stream but differ in where adapter
computations are performed. For FA, $A_\ell$ is
public and frozen while $B_{i,\ell}$ is private and
trainable. For XS, both $S_{\mathrm{in},\ell}$ and
$S_{\mathrm{out},\ell}$ are public, frozen factors,
while only $C_{i,\ell}$ is trainable. NICO and MAKI
reuse these public factors as client-independent
optical streams so that receiver-side execution does not
require storing their weight matrices at each client.

\noindent\textbf{NICO.}
NICO implements the private trainable adapter transform optically at the receiver without revealing its private weights to the server. The source encodes the shared backbone as well as the public fixed adapter factors. For XS, the noiseless receiver-side decomposes it into three-layer optical computation,
\begin{equation}
\begin{aligned}
h^{(1)}_{i,\ell}
&=S_{\mathrm{in},\ell}x_{i,\ell},\\
h^{(2)}_{i,\ell}
&=C_{i,\ell}h^{(1)}_{i,\ell},\\
\Delta y_{i,\ell}
&=S_{\mathrm{out},\ell}h^{(2)}_{i,\ell}.
\end{aligned}
\label{eq:nico_xs_computation}
\end{equation}
Here, the first and third stages use public optical
weight streams. To implement the private middle
stage, the server supplies unweighted optical
carriers, onto which the client encodes its
$r\times r$ core $C_{i,\ell}$ using local
electro-optic weight modulators. In the
digital-state implementation, private coefficients
are read from client memory and converted through
DACs and drivers, where their magnitudes control carrier
attenuation and their signs determine the balanced
detection channels. The rank-width output
$h^{(1)}_{i,\ell}$ of the first stage is detected
and used to drive an input modulator on the
locally weighted carriers. As in the
LN/S signaling of Eq.~\ref{eq:netcast_counts},
the expected differential charge in output lane
$j$ satisfies
\begin{equation}
\mathbb{E}[Q^+_{C,j}-Q^-_{C,j}]
\propto \sum_{k=1}^{r}
C_{i,\ell,jk}h^{(1)}_{i,\ell,k},
\label{eq:nico_private_core_charge}
\end{equation}
up to client-local optical normalization and
gain restoration. The resulting rank-width
$h^{(2)}_{i,\ell}$ is remodulated onto the public
$S_{\mathrm{out},\ell}$ stream to yield the final output. FA follows
similarly, using public $A_\ell$ and a locally
encoded private $B_{i,\ell}$ optical stage.
This design trades client-side modulation and
rank-width conversion for optical private-core
MACs and the blank-carrier delivery incurs additional
server-side optical supply. However, it shall be noted that no private adapter
coefficients, activations, or gradients need be
transmitted to the server.

\noindent\textbf{MAKI.}
MAKI retains the mutable adapter computation in
local electronics while executing fixed adapter
factors optically. In FA, the fixed $A_\ell$
transform is streamed optically, and the private
$B_{i,\ell}$ transform is electronic. In XS,
\emph{both} fixed transforms
$S_{\mathrm{in},\ell}$ and $S_{\mathrm{out},\ell}$
are supplied as public optical streams, with
$C_{i,\ell}$ evaluated electronically between
them. The intermediate rank signal is detected
for the electronic core and remodulated onto the
public output-factor stream.

\noindent\textit{XS rank-domain calibration and scaling.}
For each XS layer, let $Q_{1,i,\ell}$ and
$Q_{2,\ell}$ be positive diagonal rank-$r$
scales. The client calibrates $Q_{1,i,\ell}$
from local training-time rank-activation
statistics of $h^{(1)}_{i,\ell}
=S_{\mathrm{in},\ell}x_{i,\ell}$,
then freezes and persistently stores its
$r$ diagonal entries. In the current LIELLA default calibration, the server calculates
 $g_{\ell,j}$ as the inverse column $\ell_2$ norm of
$D_{\mathrm{out},\ell}^{-1}S_{\mathrm{out},\ell}$,
and the client calculates $v_{i,\ell}$ as the normalized
coordinatewise RMS of $h^{(1)}_{i,\ell}$.
With a shared, fixed $\lambda_q$, $Q_{2,\ell}$ and $Q_{1,\ell}$ are then given by
\begin{align}
    Q_{2,\ell}&=\lambda_q\operatorname{diag}(g_\ell),\\
 Q_{1,i,\ell}&=\lambda_q(\min_j g_{\ell,j})
\operatorname{diag}(v_{i,\ell}).
\end{align}
As a result, $Q_{2,\ell}$ depends only on public
fixed factors and the shared scaling setting,
whereas $Q_{1,i,\ell}$ remains local and private.
We define
\begin{equation}
\begin{aligned}
\bar C_{i,\ell}
&=Q_{2,\ell}^{-1}C_{i,\ell}Q_{1,i,\ell},\\
\bar S_{\mathrm{out},\ell}
&=D_{\mathrm{out},\ell}^{-1}
S_{\mathrm{out},\ell}Q_{2,\ell},\\
u_{i,\ell}
&=\bar C_{i,\ell}Q_{1,i,\ell}^{-1}h^{(1)}_{i,\ell}
=Q_{2,\ell}^{-1}C_{i,\ell}h^{(1)}_{i,\ell}.
\end{aligned}
\label{eq:xs_private_rank_scaling}
\end{equation}
Their product is exactly
$\bar S_{\mathrm{out},\ell}u_{i,\ell}
=D_{\mathrm{out},\ell}^{-1}
S_{\mathrm{out},\ell}C_{i,\ell}h^{(1)}_{i,\ell}$.
NICO constructs $\bar C_{i,\ell}$ locally
for its optical private core, while MAKI
evaluates $C_{i,\ell}h^{(1)}_{i,\ell}$
electronically. Both use the same public
$\bar S_{\mathrm{out},\ell}$ optical stream and 
neither sends $Q_{1,i,\ell}$ to the server.

\noindent\textbf{ELI.}
ELI performs only the shared backbone multiplication
optically, retaining all adapter transforms in
client-side electronics. This follows a similar approach
as in Opto-ViT-v2~\cite{chen2026optovit} and serves as the reference design.
This does not require an additional public distributed weight stream
shared by independent clients. However, this comes at a price of retaining all local
fixed-factor storage and electronic adapter MACs.

The optical placements require public fixed-factor
streams and, for BP, their transposed orientations
in addition to the backbone service. Their receiver
conversions, optical noise, and resource costs are
discussed in Section~\ref{sec:model}.

\subsection{Training Execution}
\label{sec:training_execution}
Figure~\ref{fig:architecture}(a) illustrates the
shared forward $W$ stream used by inference and
FO training, alongside the additional
$W^{\mathsf T}$ stream required for BP.
For \emph{forward-only (FO)} training, each client
perturbs its private adapter and evaluates losses
using only the  forward service. This reuses the public backbone and fixed-factor optical streams for NICO and MAKI without requesting their transposes. Therefore, this also provides additional privacy, as the clients who are performing training behave mostly the same as the client doing inference.

For \emph{backpropagation (BP)}, the shared service
additionally broadcasts transpose-oriented streams
for $W_\ell$ and any optically executed public adapter
factors. Each client propagates derivatives through its
private adapter and computes parameter gradients locally.
For XS, the adapter's reverse path applies
$S_{\mathrm{out},\ell}^{\mathsf T}$, $C_{i,\ell}^{\mathsf T}$, and
$S_{\mathrm{in},\ell}^{\mathsf T}$ in sequence.
BP also requires retained forward activations or recomputation through additional
forward-stream passes.

In both modes, again, the client's data and weights are kept private, as the shared source requires no client-specific information. Clients served by the common public
stream schedule can consume the same broadcasts
concurrently, while reverse requests may require additional service slots.
LIELLA, therefore, trades FO's query volume and accuracy against
BP's transpose-service requirements
 and activation memory. The evaluated optimizer recipes and their
service accounting are detailed in
Section~\ref{sec:evaluation}.

\section{Hardware and Resource Models}
\label{sec:model}
In this section, we develop a unified operation-level model that
characterizes optical receiver noise, client-side
electrical energy, and memory requirements.
By accounting for photonic computations, electronic
interfaces, and training-service requests, the model
enables us to consistently compare resource requirements across
NICO, MAKI, and ELI, under various training methods.

\subsection{Optical Noise Model}
\label{subsec:OpticalNoiseModel}
\noindent\textbf{Receiver photon budget.}
Under the LN/S signaling of
Section~\ref{sec:background}, the detected-electron
count scales with the absolute encoded weight
magnitude. Let
$\mu_M=\operatorname{mean}_{jk}|\bar M_{jk}|$
denote the mean absolute weight of a normalized
optical matrix $\bar M$, and let
$E_{\mathrm{rx},M}$ denote the average
receiver-incident optical energy per active
MAC, after any local magnitude attenuation.
The corresponding full-scale ($|\bar M_{jk}|=1$) detected-electron
count is
\begin{equation}
n_{0,M}=
\frac{\eta E_{\mathrm{rx},M}}
{(hc/\lambda)\mu_M},
\label{eq:noise_photon_budget}
\end{equation}
where $\eta$ is the detector quantum efficiency, and $hc/\lambda$ is the photon energy.

\noindent\textbf{Shot and read noise.}
We model the differential photoelectron counts
in Eq.~\ref{eq:netcast_counts} as independent
Poisson variables. Since their summed mean
is $n_{0,M}|\bar M_{jk}|$ for each active MAC,
the normalized output noise variances are
\begin{equation}
\begin{aligned}
\operatorname{Var}(\xi_{\mathrm{shot},j})
&=\frac{\sum_k(\mu_{jk}^{+}+\mu_{jk}^{-})}
{n_{0,M}^{2}}
=\frac{\|\bar M_{j,:}\|_1}{n_{0,M}},\\
\operatorname{Var}(\xi_{\mathrm{read},j})
&=\left(\frac{\sigma_e}{n_{0,M}}\right)^2.
\end{aligned}
\label{eq:noise_variances}
\end{equation}
Here, $\sigma_e$ is the RMS electronic noise
per integrated receiver read. For an active
read, the shot-noise variance depends on
the encoded weight magnitudes and is independent
of the normalized input values. We simulate shot and read noise using samples from an independent zero-mean Gaussian distribution
with the above variances. The receiver model uses $\lambda=1550$ nm, $\eta=0.8$, and $\sigma_e=127.03$ electrons RMS. And the noisy output is clamped to the receiver range before LN/S scaling restoration, with exact-zero inputs elided.

\noindent\textbf{Adapter stages and common read.}
In NICO and XS/MAKI, the final optical
adapter branch (private $B_i$ for FA or
public $S_{\mathrm{out}}$ for XS) shares a
differential receiver with the backbone
$W$. Because the two branches use
independently normalized inputs, their
contributions must be gain-matched at the
common read. Let $s_{x,i}$ denote the
backbone input normalization and $u_i$ the
local rank-coordinate input to the final
optical adapter factor
(Eq.~\ref{eq:xs_private_rank_scaling} for XS).
For each active input vector, the client sets
$t_i=\max(s_{x,i},\|u_i\|_\infty)$ and
$\rho_i=t_i/s_{x,i}\geq1$. This gain
matching is implemented through additional
receiver-side attenuation
$\eta_i=\rho_i/R\leq1$ of a fixed-headroom
public adapter stream, where $R$ bounds all
admitted $\rho_i$. The backbone photon
budget and shot noise remain unchanged,
while the adapter's received photon count
and shot noise variance depend on $\rho_i$.
For the XS common receiver, the conditional
normalized variance at output lane $j$ is
\begin{equation}
\begin{aligned}
\operatorname{Var}(\xi_{\mathrm{W+FT},j}\mid\rho_i)
={}&\frac{\|\bar W_{j,:}\|_1+
\rho_i\|\bar S_{\mathrm{out},j,:}\|_1}{n_{0,W}}+\frac{\sigma_e^2}{n_{0,W}^2}.
\end{aligned}
\label{eq:noise_common_read}
\end{equation}
\begin{figure}[t]
\centering
\includegraphics[width=\columnwidth]{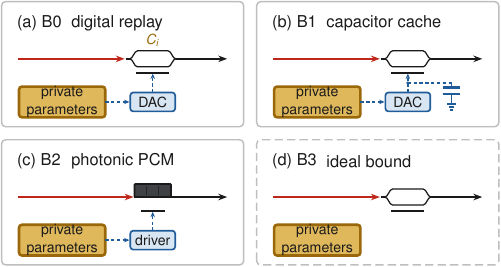}
\caption{
Illustration of private-state residency scenarios.  B0 digitally reloads private weights. B1 caches analog biases with refresh.  B2 uses nonvolatile photonic storage with programming costs. B3 assumes zero-cost delivery. }
\label{fig:residency}
\end{figure}

\begin{table}[t]
\centering
\caption{Nominal client-electrical energy
coefficients (pJ/event). B1/B2 programming
costs are conditional device estimates;
the B2 value is a lower bound.}
\label{tab:energy_coefficients}
\small
\begin{tabularx}{\columnwidth}{@{}Xr@{}}
\toprule
Event (unit) & Nominal\\
\midrule
DAC + loaded modulator (scalar input)~\cite{chen2012dac,ou2025hitop}
&2.1\\
Balanced receiver + ADC (scalar output)~\cite{zhu2020adc,sludds2022netcast}
&2.5\\
Composite rank O/E/O bridge (scalar conversion)~\cite{chen2012dac,zhu2020adc,ou2025hitop}
&5.0\\
Integrated electronic (MAC)~\cite{zimmer2019accelerator,tortorella2022redmule}
&2.9\\
Local SRAM access (byte transfer)~\cite{kneip2023sram,horowitz2014energy}
&2.5\\
B1 analog state write (bias update)~\cite{lam2026electroopticmemory}
&56\\
B2 photonic state programming (cell update)~\cite{gao2025photonicpcm}
&$\geq 4.8$\\
\bottomrule
\end{tabularx}
\end{table}

\subsection{Electrical Energy Model} 
To quantify the client-side electrical cost of shared photonic weight streaming and private adaptation, we model the streaming-and-adaptation (SA) subsystem. Its modeled operation set $\mathcal K_{\mathrm{SA}}$ covers client DAC/modulator and receiver/ADC interfaces for the shared $W$ stream (and $W^{\mathsf T}$ for BP), optical and electronic adapter operations, private-state delivery, and local optimization. We sum the corresponding electrical energy as
\begin{equation}
E_{\mathrm{SA,client}}
=\sum_{k\in\mathcal K_{\mathrm{SA}}}n_k e_k,
\label{eq:energy}
\end{equation} where $n_k$ and $e_k$ denote the execution count and electrical energy per event of category $k$, respectively. And we normalize it by served backbone MACs to get $e_W=10^{15}E_{\mathrm{SA,client}}/N_W$ in fJ per served backbone MAC. Receiver-incident optical energy is counted separately using reference optical MAC energies in Sec.~\ref{subsec:OpticalNoiseModel}. 

The energy accounting reflects the actual execution of each architecture. A shared optical $W$+adapter output incurs one common receiver/ADC read, while intermediate rank conversions are counted separately. Composite operation costs are charged only once. FO and BP costs follow their respective model-call counts, including additional forward evaluations and backward passes. We also account for runtime normalization,
private-factor generation, and associated
memory accesses. The source-equivalent model
charges FP32 divisions and per-vector SRAM
reads. Furthermore, we consider precomputed
reciprocals for fixed coefficients and per-call
coefficient reuse as conditional arithmetic
and traffic optimizations while including their additional setup and cache costs.
One-time calibration setup costs are excluded,
while runtime use of calibration coefficients
remains included.

\noindent\textbf{Private-state residency.}
To examine the energy overhead of repeated
private weight delivery and programming,
we further consider the following four private-state residency
scenarios, illustrated in Fig.~\ref{fig:residency}:
\begin{enumerate}
    \item B0 digitally replays private
weights.
    \item B1 uses a volatile capacitor-held analog cache.
    \item B2 assumes nonvolatile photonic storage.
    \item B3 idealizes zero private-delivery energy.
\end{enumerate}

The state-programming energy assumptions
are listed in Table~\ref{tab:energy_coefficients}.
B1 assumes a refresh every $\tau/100$,
where $\tau$ is the device retention time.
B2 inference setup is amortized over
100 requests. 

\noindent\textbf{Energy coefficients.}
Table~\ref{tab:energy_coefficients} lists the
nominal component costs used for the energy
results in Fig.~\ref{fig:energy}.

\begin{figure*}[!t]
\centering
\definecolor{liellaNICO}{HTML}{2E2B35}
\definecolor{liellaMAKI}{HTML}{B34B5C}
\definecolor{liellaELI}{HTML}{C6A047}
\includegraphics[width=\textwidth]{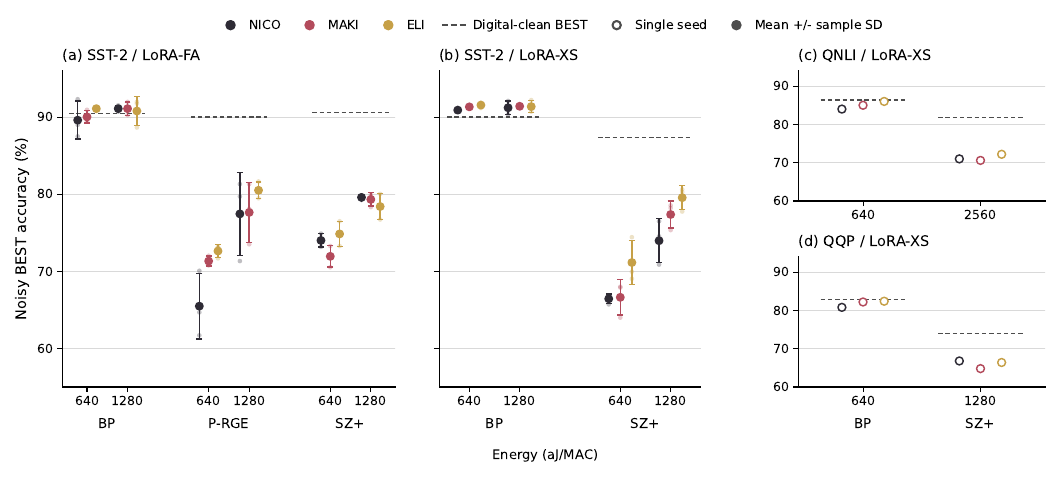}
\caption{\textbf{Private adaptation under photonic receiver noise}.
(a,b) SST-2 LoRA-FA and LoRA-XS accuracy under BP and
forward-only optimization at 640 and 1280~aJ per MAC.
(c) QNLI LoRA-XS with BP at 640~aJ and SubZero+ at 2560~aJ.
(d) QQP LoRA-XS with BP at 640~aJ and SubZero+ at 1280~aJ.
SST-2 uses 872 validation examples; QNLI and QQP each use
a separate fixed 500-example validation panel.
Colors denote NICO, MAKI, and ELI. Error bars show sample
standard deviations across accepted hardware-noise seeds;
open markers denote a single seed. Gray dashed lines indicate
task- and method-matched, separately digital-clean-trained
\BEST{} references.}
\label{fig:learning}
\end{figure*}

\subsection{Client Memory and Storage Models}
\noindent\textbf{Peak client memory.} We estimate it by accounting for simultaneously live tensors across execution phases, including persistent state and transient workspace. Shared allocations are counted once, and the maximum memory footprint across phases determines the actual peak memory. We assume the inputs are pre-embedded in FP32 for simplicity, which gives $X\in\mathbb R^{B\times S\times2048}$ (0.5625 MiB at $B=1,S=72$ with $B$ being the batch size). Therefore, embedding lookup is excluded as its public weights can also be streamed for client-local computation, avoiding persistent storage of the 250 MiB table without disclosing private token IDs. Clients retain 45 frozen FP32 RMSNorm vectors (0.3515625 MiB), private state, and transient attention/layer workspace. Autoregressive KV caching is disabled, and workspace is reused across layers.  We also account for attention, intermediate activations, and loss computation in addition to private adapter storage. In particular, full-vocabulary cross-entropy can dominate training memory, while selecting supervised positions or using fused kernels can reduce its workspace~\cite{wijmans2025cce}. We distinguish these potential optimizations from the implemented simulation when comparing inference, FO, and BP.

\noindent\textbf{Persistent state.} For FA, the private $B_{i,\ell}$ factors require 24.06 MiB, plus 9.63 KiB for client-private rank calibration $Q_{1,i,\ell}$. For XS, the private cores $C_{i,\ell}$ and calibration $Q_{1,i,\ell}$ require 38.5 KiB and 4.81 KiB, respectively, totaling 43.31 KiB. Both estimates assume 154 adapted layers and FP32 precision.

\begin{table}[t]
\centering
\caption{Private-state capacity in the B0-B2 model.}
\label{tab:private_bank_capacity}
\setlength{\tabcolsep}{3pt}
\begin{tabularx}{\columnwidth}{@{}Xrrr@{}}
\toprule
Resource & FA16 & FA8 & XS8\\
\midrule
Private parameters & 6,307,840 & 3,153,920 & 9,856\\
B1: capacitor bank & 90,112 & 45,056 & 64\\
B2: photonic PCM cells & 6,307,840 & 3,153,920 & 9,856\\
\bottomrule
\end{tabularx}
\end{table}

\noindent\textbf{Physical capacity assumptions.}
Table~\ref{tab:private_bank_capacity}
reflects a device-density and retention tradeoff.
B1 uses relatively large, volatile capacitors to
hold analog modulator biases. Therefore, we reuse one
largest-module bank, charging reprogramming
and refresh. B2 uses much denser, nonvolatile photonic PCM
cells, making full-model residency plausible
without inter-module reloads, while charging
initial programming and training updates.
From the reference devices
~\cite{lam2026electroopticmemory,gao2025photonicpcm},
the B1 reference offers 5.33--5.97 effective bits,
while B2 reports 209 levels (about 7.71 bits)
and a 6,000-cycle single-cell endurance point.
Our capacity and energy model assumes one capacitor-held analog state per B1 coefficient or one nonvolatile photonic cell per B2 coefficient; neither implies FP32 optical precision. FP32 masters and calibration remain digital. Simple quantization (XS, range rescaled for FA) to the corresponding level changes noisy accuracy by less than 1 percentage point across tested energies in fixed-checkpoint inference.

\section{Experimental Setup}
\label{sec:evaluation}

\subsection{Model and Workloads}
LIELLA is evaluated using FP16 1.1 B TinyLlama,
a 22-layer Transformer with a hidden
dimension of 2048 and a vocabulary of
32,000 tokens~\cite{zhang2024tinyllama}.
Three GLUE (SST-2, QNLI, QQP~\cite{socher2013sst,wang2018glue}) tasks are being evaluated in this work. For SST-2, we use 1,000 training examples
and the full 872-example validation set.
The QNLI and QQP evaluation panels
contain 500 each, respectively.  Given the computational cost of repeated
training under simulated hardware noise,
we use a common training seed (42) to
facilitate controlled comparisons across
hardware configurations.

\subsection{Adaptation and Optimization Configurations}
We evaluate backpropagation (BP) for both FA and XS.
For forward-only (FO) adaptation, FA uses
MobiZO's parallelized randomized gradient estimation
(P-RGE)~\cite{gao2025mobizo} and
SubZero+~\cite{yu2026subzeroplus}, while XS uses
SubZero+ in the main noisy-training comparisons.
P-RGE uses four antithetic direction pairs
(eight loss evaluations per update) and SGD.
SubZero+ uses 99 one-sided directions plus a
shared baseline, projected-subspace Adam,
and a basis refresh every 50 updates.
We adopt representative batch settings from the
reference methods: four examples per P-RGE
evaluation and a batch size of 16 for SubZero+.
For FA, SubZero+ updates the full private
factor through a rank-eight optimization
subspace without changing the adapter
parameterization. For each optimization method, we conduct a limited pilot hyperparameter search to select its configuration. Training is generally continued until validation performance stabilizes, except for experiments with a predefined update budget.

\subsection{Evaluation Protocol and Metrics}

The validation accuracy is reported at the best
checkpoint (\BEST, the earliest
update in case of ties).
Noise-OFF (\OFF) evaluates the same checkpoint with receiver
noise disabled, without retraining.
Hardware-noise seeds are aggregated only
within matched configurations. Accuracy is evaluated through 
simulations incorporating optical shot and
read noise. Electrical energy and client
memory are estimated using the models
in Section~\ref{sec:model}.
Resource estimates use a reference sequence
length of $S=72$, while training retains
the actual padded sequence lengths.
With a full-position vocabulary head,
each sequence involves
$N_W\sim74\,\mathrm{B}$ backbone MACs.
We report inference electrical energy per
backbone MAC and training energy per update,
accounting for each method's actual batch
size and query count.

\section{Results}
\label{sec:results}

\begin{figure*}[!t]
\centering
\includegraphics[width=\textwidth]{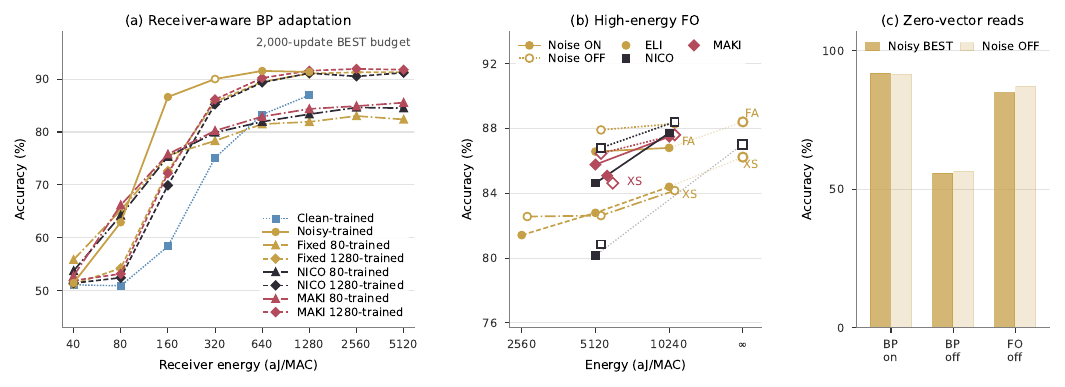}
\caption{\textbf{Low-Energy BP Adaptation, High-Energy FO Recovery, and Zero-Vector Gate Sensitivity.}
(a) XS--BP noisy-inference curves within 2,000-update budget under various energy.  The ELI noisy-trained curve averages three independently trained checkpoints, whereas each fixed-energy curve averages three noisy inference draws of one selected frozen checkpoint. (b) High-energy FO accuracy. Noise-ON/OFF readouts use
distinct line styles, with ELI circles, NICO squares, and MAKI diamonds.
The XS/NICO noise-OFF point at 5120~aJ/MAC is 80.8\%.
Independent noiseless-trained references for FA/ELI,
XS/ELI, and XS/NICO appear at conceptual $\infty$,
joined only by visual reference guides, not matched checkpoints.
(c) FA/ELI zero-vector gating: BP ON/OFF and ungated
P-RGE FO noisy and noise-off accuracy under \BEST{} checkpoint.}
\label{fig:hardware_adaptation}
\end{figure*}

\subsection{Learning under Photonic Noise}

\noindent\textbf{Private adaptation is feasible
under low receiver photon budgets.}
Figure~\ref{fig:learning} shows that both BP and
FO support independent private fine-tuning over
the same frozen photonic backbone under receiver
noise, with broadly comparable SST-2 accuracy
across NICO, MAKI, and ELI. At 640~aJ per active
optical MAC, BP reaches up to 91.1\% and 91.6\%
on SST-2 with FA and XS, respectively, and
84.0--86.0\% on QNLI and 80.8--82.4\% on QQP
with XS, the latter approaching its 82.8\%
digital-clean reference. SubZero+ also supports noisy QNLI adaptation
across all three adapters at 2560~aJ and QQP
adaptation at 1280~aJ in the same placements. 

\noindent\textbf{SubZero+ improves low-energy FO accuracy
while spending more forward-query.}
At 640~aJ, FA--SubZero+ achieves higher mean
accuracy than P-RGE across all three architectures,
with the largest gain in NICO
(74.0\% versus 65.5\%).
Each SubZero+ gradient estimate uses 100
forward evaluations, equivalent to 12.5 P-RGE updates. Across nine  trajectories per method
at 640~aJ, the selected \BEST{} checkpoints
require approximately 44k versus 13k estimator
forward evaluations per run on average (3.3$\times$).

\noindent\textbf{BP is substantially more robust
under photonic noise.}
At 640~aJ, both FA--BP and XS--BP already
achieve accuracy close to their digital-clean
references. In contrast, even at 1280~aJ,
FA--P-RGE and XS--SubZero+ remain approximately
9.5 and 7.8 percentage points below their
respective clean references in ELI.
These results motivate the development of
more noise-robust FO optimization methods.
For XS, we focus on SubZero+ because clean
pilot experiments showed that P-RGE required
substantially more forward evaluations to
reach comparable accuracy while still
falling short of BP.

\subsection{Hardware-Aware Adaptation}

\noindent\textbf{Noisy-trained BP outperforms clean-trained BP
at low receiver energies.}
In Figure~\ref{fig:hardware_adaptation} (a), we plot the 
independently noisy-trained XS--BP/ELI \BEST{}
checkpoints against a fixed, digitally clean-trained
checkpoint, evaluating them under receiver noise.
At 80 and 160~aJ/MAC, noisy-trained BP achieves
mean accuracies of 62.9\% and 86.6\%, respectively,
compared with 50.9\% and 58.4\% for the
clean-trained reference. The high-energy \BEST{} checkpoints at 640 and 1280 aJ/MAC
reach 91.6\% and 91.4\%, compared to 83.3\% and
87.0\% for the same clean-trained reference.
These gains suggest the potential for receiver-specific
hardware adaptation through private PEFT, consistent
with prior photonic work on low-rank hardware-error
compensation and noise-resilient fine-tuning~
\cite{huang2024opticalLoraCalibration,
taniguchi2025alora,chen2026optovit}.

\begin{figure*}[t]
\centering
\includegraphics[width=.96\textwidth]{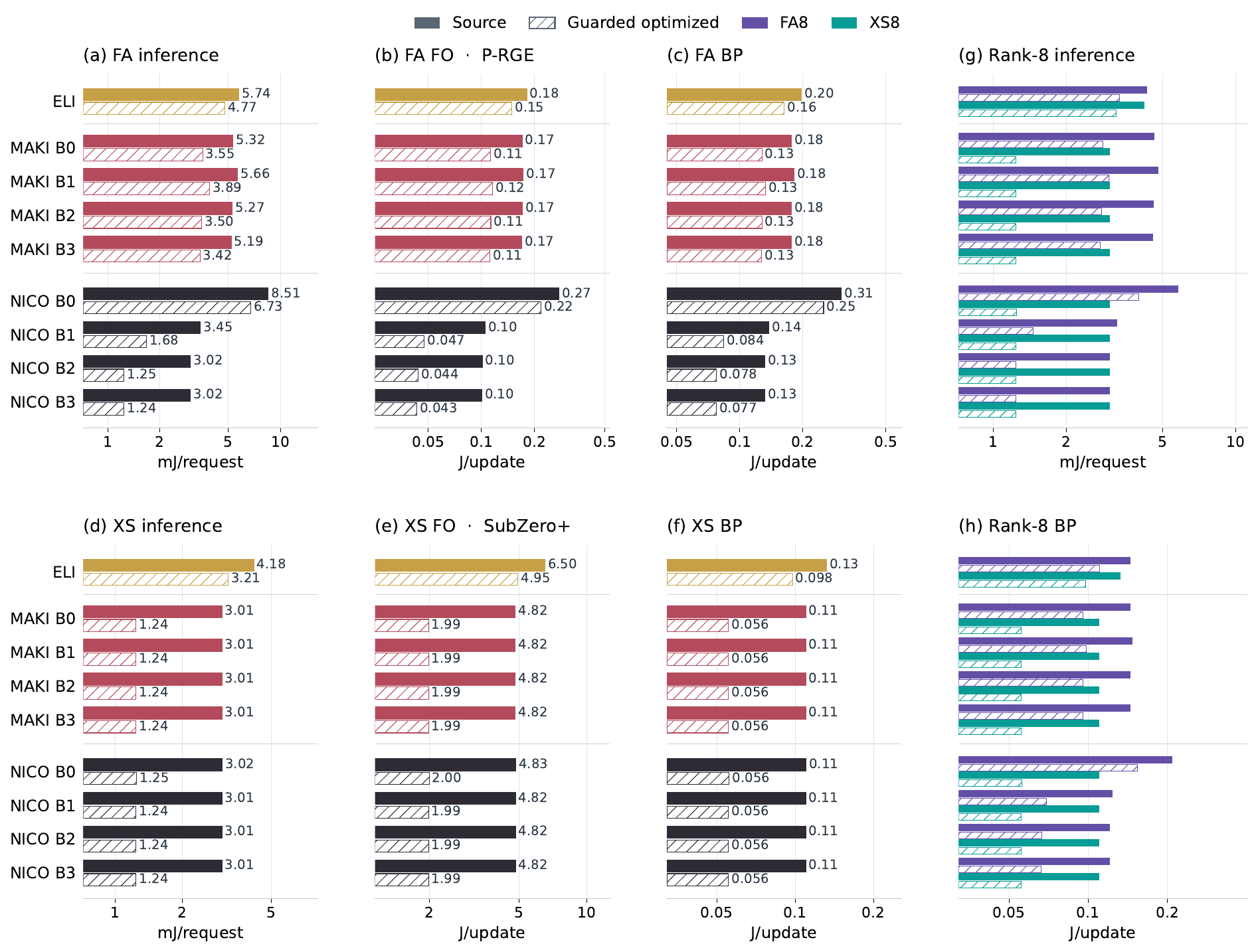}
\caption{\textbf{Client SA electrical energy.}
(a)--(f) FA (top) and XS (bottom) inference,
FO, and BP; (g,h) rank-matched FA8/XS8
inference and BP. Each panel compares ELI
with MAKI/NICO B0--B3. Solid/hatched bars
denote the simulated-source/guarded-optimized arithmetic.}
\label{fig:energy}
\end{figure*}

\noindent\textbf{Client-specific adaptation accommodates
heterogeneous optical receivers.}
Figure~\ref{fig:hardware_adaptation}(a) further shows
frozen XS--BP trained at
80 and 1280~aJ/MAC under three different adapters, with checkpoints selected
within a common 2,000-update \BEST{} budget
and evaluated at eight receiver energies.
At 80~aJ/MAC, the low-energy-trained ELI adapter
achieves 65.2\% versus 54.2\%, while at
1280~aJ/MAC, the high-energy-trained adapter
achieves 91.1\% versus 81.9\%.
In both cases, receiver-matched training yields
over 11\% relative accuracy improvement
over the mismatched counterpart and similar conclusions hold for MAKI and NICO.
In a shared optical broadcasting system,
propagation loss and coupling efficiency can
provide different clients with distinct receiver photon
budgets. These results motivate independent,
receiver-aware private fine-tuning for each client
rather than relying on a single shared adapter.

\noindent\textbf{FO approaches noiseless accuracy
at high receiver energies.}
Figure~\ref{fig:hardware_adaptation}(b) compares
high-energy noisy FO \BEST{} accuracy with
same-checkpoint noise-OFF readouts and
separately noiseless-trained ELI references.
At 5120~aJ/MAC, FA--P-RGE and XS--SubZero+
in ELI achieve noisy accuracies of 86.6\%
and 82.8\%, while disabling inference noise
on those same checkpoints yields 87.9\%
and 82.6\%, respectively.
At 10240~aJ/MAC, the noisy accuracies
reach 86.8\% and 84.4\%, with same-checkpoint
noise-OFF readouts of 88.3\% and 84.2\%.
Both noisy results lie within 2 percentage
points of the separately noiseless-trained
references of 88.4\% and 86.2\%.
Occasional small decreases in XS noise-OFF accuracy are likely due to statistical fluctuations.

\noindent\textbf{BP's accuracy advantage entails
additional hardware and trust requirements.}
Figure~\ref{fig:hardware_adaptation}(c) shows
that zero-vector gating raises FA--BP accuracy
from 55.7\% to 91.9\% at 1280~aJ/MAC.
In a separate fixed-state diagnostic, nearly
65\% of transpose inputs were exactly zero
and elided, which forced their physical reads
to introduce receiver noise that propagated
into upstream error signals.
Moreover, BP requires an additional
$W^{\mathsf T}$ stream, which an untrusted
server could withhold to suspend training
or alter to corrupt client-side gradients,
as illustrated by related server-side
manipulation risks in split learning~
\cite{yu2024chronicPoisoning}.
In contrast, ungated P-RGE FO reaches 84.9\%
using only the forward stream.
These hardware and trust dependencies
further motivate our preference on more accurate, query-efficient
FO methods.

\begin{figure*}[t]
\centering
\includegraphics[width=\textwidth]{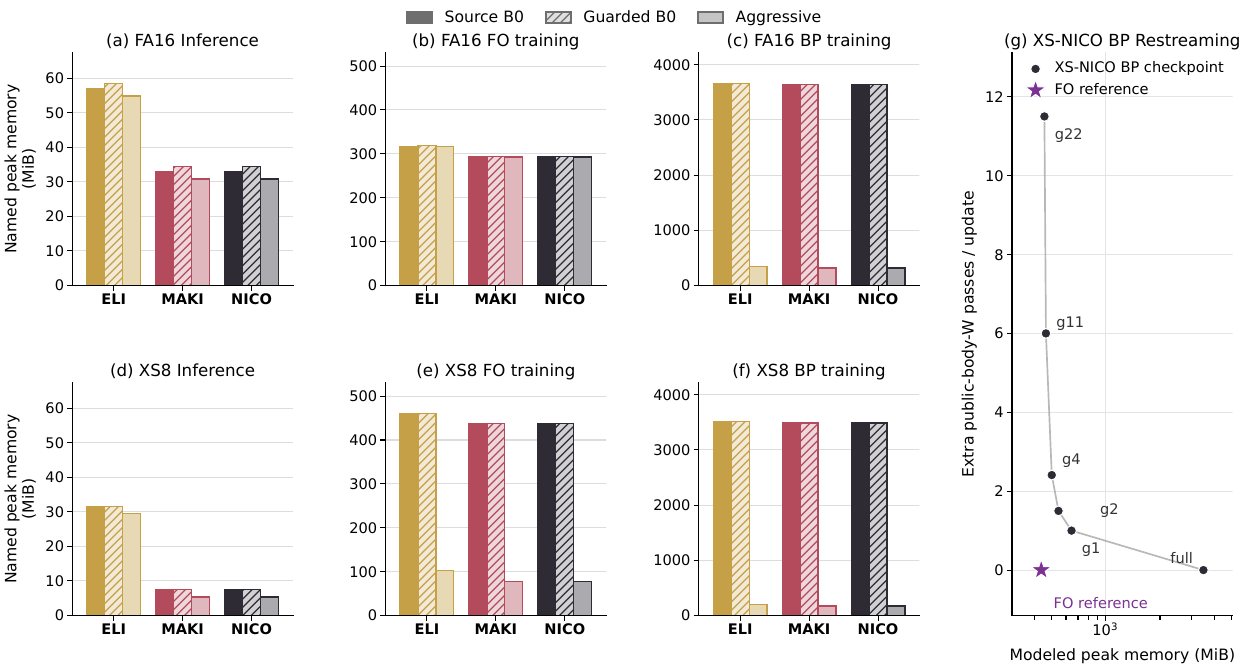}
\caption{\textbf{Client-side memory.}
(a)--(f) Peak named-buffer memory for FA and XS
under ELI, MAKI, and NICO, comparing source,
guarded arithmetic, and memory-only aggressive optimization projection. MAKI and NICO use B0
digital replay. All use $S=72$; inference uses
(Batch)$B$$=1$, FA P-RGE FO $B=4$, and XS SubZero+ FO
and BP $B=16$.
(g) Checkpoint-only XS-NICO BP memory versus additional backbone
restreaming; the purple star marks the unoptimized
XS FO reference. "g" denotes group size, the number of Transformer blocks per checkpoint segment, while “full” denotes full activation caching. }
\label{fig:memory}
\end{figure*}

\subsection{Client-Side Electrical Energy}

\noindent\textbf{MAKI reduces electronic computation.}
By offloading fixed adapter transforms to
the optical stream, MAKI eliminates their
local electronic MACs and weight accesses
while keeping private computation electronic.
Figure~\ref{fig:energy} shows that, even under
B0, this reduces energy relative to ELI
by 7.4\% for FA inference and 10.7\% for
FA BP under the baseline model.
For XS inference, the saving further reaches 28.0\%
as both fixed factors are offloaded.

\begingroup\emergencystretch=2em
\noindent\textbf{Private-weight residency benefits
NICO far more than MAKI.}
Although FA-NICO performs private-$B$ MACs
optically, B0 repeatedly fetches and drives
6.3 million private coefficients through
SRAM and DAC/modulators, raising inference
energy to 8.51~mJ/request, 48.1\% above ELI.
Volatile analog caching (B1) reduces this
to 3.45~mJ, while nonvolatile residency
(B2) further lowers it to 3.02~mJ,
saving 40.0\% and 47.5\% over ELI,
respectively.
In contrast, FA-MAKI gains little from residency, with B1 even increasing energy by 6.5\% over B0.
Because MAKI retains electronic private-$B$
MACs, analog residency introduces
additional readout, conversion, and
staging costs without eliminating
these computations.\par\endgroup

\noindent\textbf{Arithmetic optimization reveals
larger optical savings.}
Further energy breakdown shows that repeated
normalization and rescaling substantially
offset the benefits of optical computation,
particularly in MAKI and NICO.
We therefore consider the optimization by folding fixed scaling factors
and reusing cached reciprocals across
normalization operations, with numerical checks and a conditional fallback to avoid extremely small numbers.
These optimizations benefit MAKI and NICO
more than ELI, whose electronic private-core
MAC costs remain largely unaffected.
Under the same guarded arithmetic policy,
FA-MAKI B0 saves 25.6\% over ELI, while
FA-NICO B1/B2 achieves 64.9/73.9\% savings.
FA BP exhibits the same trend, with
21.8\% savings for MAKI B0 and
48.7/52.5\% for NICO B1/B2.
For XS inference, MAKI B0 achieves
61.3\% savings even without optical
private-weight residency.

\noindent\textbf{Compact XS private cores
mitigate residency overhead.}
XS has only 9,856 private $C$ coefficients,
compared with 6.3 million in FA16,
greatly reducing private-weight delivery.
Under guarded B0, XS-NICO consumes
just 1.25~mJ/request, 61.1\% below XS-ELI.
Its energy varies by less than 0.7\%
across B0--B3, while XS-MAKI achieves
nearly identical energy without an
optical private core.
To isolate the effect of adapter rank,
Figure~\ref{fig:energy}(g)--(h) further
compares FA8 and XS.
Under guarded NICO B0, XS reduces
inference energy by 68.6\%, although this advantage narrows to 14.7\% under
B1 and nearly vanishes under B2, while XS still saves 16.5\% in
B2 BP.
These findings provide a new hardware-driven motivation for developing more efficient PEFT methods that unlock photonic energy savings without relying on costly private-weight residency.

\subsection{Client Memory and Private Storage}

\noindent\textbf{XS reduces inference memory beyond
private storage.}
Under B0, XS-NICO requires only 7.46~MiB of client inference memory, compared with 32.87~MiB for FA16-NICO, a 4.4$\times$ reduction. With MLP chunking~\cite{kitaev2020reformer},
this falls further to 5.35~MiB. Notably, both footprints are smaller than the 24.06~MiB required to store FA16's private weights alone. However, ELI does not have such an advantage, as it needs to store the fixed adapter factors, while XS-MAKI achieves the same 7.46 MiB as XS-NICO. Although shared execution buffers dilute the 640$\times$ private-parameter reduction, XS and photonic streaming jointly enable an extremely low runtime memory budget.

\noindent\textbf{Training memory can be provisioned separately.}
Training workspace largely eclipses adapter-size
savings where baseline XS-NICO FO requires
436.79~MiB, while full-cache BP requires
3486.82~MiB, about 1.7$\times$ the FP16
backbone weight. However, additional memory could be provisioned
only during adaptation, reused for multiple devices training and detached for deployment. In addition, we examine established memory-saving strategies, including MLP chunking~\cite{kitaev2020reformer}, selective supervised-token head computation related to memory-efficient full-vocabulary loss~\cite{wijmans2025cce}, and activation recomputation~\cite{chen2016sublinear}. For XS-NICO under guarded arithmetic, these optimizations reduce FO memory to 77.85~MiB named (79.09~MiB bounded SRAM), approximately 27$\times$ smaller than the FP16 TinyLlama weights. FA-FO benefits less because estimator-return state dominates its memory footprint.

\noindent\textbf{BP trades activation memory
for backbone restreaming.}
BP must retain intermediate activations
for backward computation, making activation
storage a major memory bottleneck.
To reduce this overhead, we consider
activation checkpointing~\cite{chen2016sublinear},
which discards intermediate activations
and recomputes them by restreaming the backbone.
For XS-NICO, one additional full-body $W$
pass per update reduces memory from
3486.82 to 643.93~MiB.
Increasing restreaming to 11.5 passes
further reduces memory to roughly 450~MiB
before the additional head and MLP optimizations.
With all three modeled optimizations, guarded BP reaches
170.19~MiB named (180.02~MiB bounded SRAM) and requires 2.28$\times$ the bounded
memory of optimized XS-FO, alongside transposed-weight streaming and additional
recomputation costs. This composition adds 88 all-token gate/up matrix
forward evaluations beyond g22's 253 extra body-block forwards;
panel~(g) isolates checkpointing alone.
These requirements further motivate more
accurate, query-efficient, and noise-robust
FO methods.

\section{Discussion and Related Work}
Prior work has explored broadcast inference (Netcast and AIR-LLM)
~\cite{sludds2022netcast,hamerly2022netcast,gao2026airllm},
photonic and analog acceleration or learning (Albireo, FICONN,
L2ight, Mirage, Opto-ViT-v2, ALoRA, and AHWA-LoRA)
~\cite{shiflett2021albireo,bandyopadhyay2024ficonn,
gu2021l2ight,demirkiran2024mirage,chen2026optovit,
taniguchi2025alora,li2026ahwalora},
and distributed or edge fine-tuning (Petals, DLoRA, MobiZO,
and MeBP)~\cite{borzunov2023petals,gao2024dlora,
gao2025mobizo,song2025mebp}.
LIELLA builds on these directions by separating shared,
client-independent weight delivery from local, mutable
private adaptation, without per-client server-side backbone
computation. It can be further viewed as a general architecture, which is not only limited to the Netcast style service, but can also be extended to AIR-LLM-style wireless RF weight broadcasting or QAMNet's coherent in-phase/quadrature optical computation ~\cite{bacvanski2024qamnet}, which could naturally preserve the public/private separation with different receivers, calibration and noise models. Related work further demonstrates the potential for strong security guarantees in optical computing through quantum-secure multiparty learning~\cite{sulimany2025quantumsecure}.

The evaluated LoRA, LoRA-FA, and LoRA-XS variants
~\cite{hu2022lora,zhang2023lorafa,balazy2024loraxs}
instantiate rather than define LIELLA's adapter interface.
FourierFT uses sparse trainable spectral coefficients
~\cite{gao2024fourierft}, suggesting optical Fourier
processing for client-specific adapters.
Quantum-PEFT uses a compact Pauli-based unitary
parameterization~\cite{koikeakino2025quantumpeft},
suggesting programmable photonic quantum circuits
as another potential private computation path. These are all compatible with light-based computation and motivate further optical computation favorable PEFT development. 

Forward-only and gradient-free methods span MeZO,
SubZero, and SubZero+~\cite{malladi2023mezo,
yu2025subzero,yu2026subzeroplus};
variance-reduced or adaptive ZO methods such as
MeZO-SVRG, ZO-AdaMU, HiZOO, PaZO, and
DiZO~\cite{gautam2024mezosvrg,jiang2024zoadamu,
zhao2025hizoo,zhao2025pazo,tan2025dizo};
and structured or batched search in LOZO,
AdaZeta, EGGROLL, and
FZOO~\cite{chen2025lozo,yang2024adazeta,
sarkar2026eggroll,dang2026fzoo}.
These ZO methods can offer further security and memory benefits compared to the BP, however, they may require many noisy forward evaluations and are less noise robust.
More fundamentally, Transformers~\cite{vaswani2017attention} are not particularly friendly, as attention and training-time activation buffers still impose substantial client memory demands, which may imply a demand for better optical-native model architecture for LLM.

Together, we anticipate the hardware realization of LIELLA, integrating shared optical weight streaming with client-side private adapters. Such implementations would enable end-to-end
characterization of energy, latency, and scalability,
while opening new opportunities for photonic-native
model architectures, hardware-efficient PEFT,
and noise-robust FO optimization.

\section{Conclusion}
We presented \liella, enabling independent private
adaptation across clients sharing a frozen photonic
weight stream. Through photonic adapter co-design
and optical-electronic placement, \liella{} significantly reduces
client-side storage and electrical costs.
We further characterize forward-only and
backpropagation services under photonic noise,
revealing their trade-offs in accuracy, energy,
and memory. Beyond the architecture, our findings
provide new hardware-driven motivations to develop more compact and effective PEFT methods, as well as more accurate, query-efficient,
and noise-robust ZO optimization algorithms.

\section{Acknowledgments}
We thank Allen Zang, and Junming Gong for their helpful discussions. We acknowledge support from the ARO(W911NF-23-1-0077), ARO MURI (W911NF-21-1-0325), AFOSR MURI (FA9550-21-1-0209, FA9550-23-1-0338), ONR MURI (N000142612102), NSF (OSI-2326767, OSI-2426975, EEC-2550064,OSI-2553578, OSI-2553619).
This work was also completed with resources provided by the University of Chicago’s Research Computing Center. We acknowledge NVIDIA for supporting the GPUs used in this work through the NVIDIA Academic Grant Program. We also acknowledge the use of Google Cloud Platform (GCP) computing resources in this work. K.S. acknowledges support from the European Union’s Horizon Europe research and innovation programme under the Marie Skłodowska-Curie grant agreement No. 101202109.

\section{Generative AI Disclosure.}
The research ideas, system design, experimental methodology, and scientific interpretations were developed and directed by the human authors. Generative AI tools, including OpenAI ChatGPT and Codex, assisted with code implementation, data organization, and manuscript editing under the authors' supervision. All technical content, experimental results, and conclusions were reviewed and validated by the authors, who take full responsibility for the work.

\bibliographystyle{ACM-Reference-Format}
\bibliography{references}
\end{document}